\documentclass{optica-article}

\journal{opticajournal} % for journals or Optica Open

\articletype{Research Article}

\usepackage{lineno}
\begin{document}

\title{Quantum Key Distribution in the Iberian Peninsula}

\author{Vicky Domínguez Tubío\authormark{1,2,*,$^\dagger$}, Mario Badás Aldecocea\authormark{3,$^\dagger$}, David L. Bakker\authormark{4}, Gustavo C. Amaral\authormark{4}, Diego López\authormark{5} and Johannes Borregaard\authormark{6} }
\address{\authormark{1}QuTech, Delft University of Technology, 2628 CJ Delft, The Netherlands\\
\authormark{2}Kavli Institute of Nanoscience, Delft University of Technology, 2628 CJ, Delft, The Netherlands\\
\authormark{3}Space Engineering Department, Delft University of Technology, 2629 HS Delft, The Netherlands\\
\authormark{4}Quantum Technology Department, The Netherlands Organization for Applied Scientific Research, 2628 CK Delft, The Netherlands\\
\authormark{5}Telefónica gCTIO/I+D, Madrid, Spain\\
\authormark{6}Department of Physics, Harvard University, Cambridge, Massachusetts 02138, USA\\
\authormark{\dag}These authors contributed equally to this work.\\
\authormark{*}vicky.dotu@gmail.com}

\begin{abstract*} 
A promising use of quantum networking is quantum key distribution (QKD), which can provide information-theoretic security unattainable by classical means. While optical fiber-based QKD networks suffer from exponential loss, satellite-assisted quantum communication offers a scalable solution for long-distance secure key exchange. In this work, we propose and evaluate a satellite-based QKD setup covering the Iberian Peninsula, linking Madrid with Barcelona, Bilbao, and Lisbon. Our proposed setup uses a Low-Earth-Orbit (LEO) state-of-the-art satellite equipped with a spontaneous parametric down-conversion (SPDC) source to distribute entangled photon pairs to ground stations. Considering vibrations in the satellite, we optimize the beam waist to enhance the transmission probability and improve the secret key rate (SKR). Our results show that key rates sufficient for real-world applications, such as secure communication between hospitals, using hybrid classical-quantum protocols are feasible with existing protocols.  Our results highlight the viability of near-term satellite-based QKD networks for national-scale secure communications.

\end{abstract*}

%%%%%%%%%%%%%%%%%%%%%%%%%%  body  %%%%%%%%%%%%%%%%%%%%%%%%%%
\section{Introduction}

The realization of a quantum internet holds the potential to revolutionize multiple fields, including secure communication \cite{BENNETT20147,Ekert_91,Pirandola2020}, high-precision sensing networks \cite{Komar2014,Guo2020,Liu2021}, and distributed quantum computing \cite{buhrman_distributed_2003}. In particular, quantum key distribution (QKD) enables communication with information-theoretic security, a level of protection unattainable by classical methods. 

Quantum communication relies on photons to carry quantum information. However, their transmission is limited by loss and, unlike classical signals, quantum signals cannot be amplified using conventional techniques due to the no-cloning theorem \cite{Wootters1982}. To mitigate this, quantum repeater architectures have been proposed. These divide the total distance into shorter segments, where quantum information can be transmitted directly. The segments are then connected using quantum teleportation \cite{Duan2001,Sangouard2011} or quantum error correction \cite{Muralidharan2014,Munro2012} at the repeater nodes, enabling faithful transmission over the total distance. In fiber-based systems, however, signal attenuation increases exponentially with distance, requiring hundreds of repeater nodes for continental-scale coverage.

Satellite-based free-space QKD offers a promising alternative for long-distance communication without the need for complex repeater infrastructures. Notably, the Micius satellite has demonstrated QKD over distances from 500 to 7,000 km using a trusted-node architecture \cite{qkd_trusted_node,vienna-beiging-qkd}. More recently, entanglement-based QKD was successfully achieved without relying on a trusted node, decreasing the risk of eavesdroppers, reaching up to 1,200 km \cite{yin_entanglement-based_2020,yin_satellite-based_2017}—distances well beyond the capability of terrestrial fiber networks.

%For optical fiber-based quantum repeaters, the transmission between the repeater nodes decreases exponentially with distance, and hundreds of repeater nodes are required to cover distances at the continental scale. Alternatively, for long-distance links, satellite-assisted free space links are a promising near-term approach that circumvents the need of complex quantum repeaters to compensate transmission loss. For example, QKD has already been demonstrated with the Micius satellite performing as a trusted nodes for distances ranging from 500 to 7000 km \cite{qkd_trusted_node,vienna-beiging-qkd}. However, assuming that the satellite is a trusted node poses a security concern. In a more advanced approach, entanglement-based QKD was achieved without relying on a trusted node, spanning distances of up to 1,200 km \cite{yin_entanglement-based_2020,yin_satellite-based_2017}, which is well outside the reach of any current fiber-based approach.

Fiber-based QKD has also made significant progress, enabling high-speed, secure communication across metropolitan areas. Recent demonstrations include the protection of multiple VPNs \cite{VPN_JPMorgan} and secure data exchange between hospitals in the Madrid area via a trusted-node QKD network \cite{telefonica_vithas_qkd_2025}. The next logical step is to extend these capabilities from city-scale to national-scale coverage.

%Other recent works have also demonstrated fiber-based QKD \cite{2024_Martin,VPN_JPMorgan}, where they have shown that they can use QKD to secure multiple independent, high-speed virtual private networks (VPNs) \cite{VPN_JPMorgan}, or that they can connect two hospitals using QKD \cite{telefonica2025blindaje}. In the latter, the two hospitals are in the Madrid Metropolitan area and use a QKD network based on trusted nodes. The next step is to move from metropolitan to national distances. To do so, we propose a setup where a space-based link distributes key between different cities in the Iberian Peninsula- Madrid, Barcelona, Bilbao and Lisbon, as shown in Fig.\ref{fig:fig_1}(a).

To this end, we propose a satellite-based QKD architecture to interconnect major cities across the Iberian Peninsula— Madrid with Barcelona, Bilbao, and Lisbon—as shown in Fig. \ref{fig:fig_1}(a). 
Our system employs a satellite like Micius in a Low-Earth-Orbit (LEO) equipped with a spontaneous parametric down-conversion (SPDC) source to distribute entangled photon pairs to multiple ground stations, implementing the BBM92 protocol \cite{yin_satellite-based_2017}. In our setup, we consider the vibrations in the satellite and propose a novel optimization of the beam waist such that it suppresses the effect of the pointing jitter and consequently it increases the transmission probability to the ground stations. %This beam optimization comes with finding the best beam waist value for the size of the telescope, accounting for truncation and jitter effects while the telescope is at the line of sight of the ground stations.

%The satellite is equipped with an onboard spontaneous parametric down conversion (SPDC) source that fires entangled photons to the ground stations, ensuring end-to-end quantum communication. We carry out the BBM92 protocol previously demonstrated with the Micius satellite \cite{yin_satellite-based_2017}. 

%We compare the performance of the satellite with and without the optimization, showing that the optimization increases the secret key rate (SKR) by a factor of... We see that a single Low-Earth-Orbit (LEO) satellite link covering the Iberian Peninsula is enough to generate the SKR needed to perform the current use case already shown in the mad-qci of encrypting the communication between two hospitals with QKD. This protocol only requires the first layer to be quantum, generating the key expansion using an internet protocol, IPsec. Going a step beyond, we also check if the raw key provided by the satellite allows us to implement Virtual Private Network (VPN) services using the ETSI-QKD-014 protocol \cite{VPN_JPMorgan}. Both these use cases are examples of hybrid protocols, where the first layer of the network is quantum and then the management and renewal of the keys is classical (that is where the ETSI protocol comes into play).

We evaluate the system's performance with and without optimization, demonstrating that the latter significantly enhances the secret key rate (SKR)—by a factor of approximately 10. In our model we also consider atmospheric conditions in real time, following the work done in \cite{Gustavo_paper}. A single satellite link proves sufficient to support secure communication across the Iberian Peninsula. Our results show that the SKR is adequate to support existing QKD use cases such as hospital-to-hospital encryption demonstrated in the MAD-QCI project, using hybrid protocols where the quantum layer handles key generation and classical layers manage key expansion via IPsec. Furthermore, we check that for more demanding use cases such as supporting VPN services using the ETSI-QKD-014 protocol \cite{VPN_JPMorgan}, we would need to improve the entangled photon pair source rate by 3 order of magnitude, i.e. 1 GHz. While such a high rate has not currently been demonstrated on a satellite, it is compatible with performances demonstrated in laboratory environments simulating LEO satellite conditions \cite{merolla2022high}.

%assess the feasibility of supporting VPN services using the ETSI-QKD-014 protocol \cite{VPN_JPMorgan}. These findings establish that satellite-optimized QKD is not only feasible but also immediately applicable to current and future quantum-secured network infrastructures at the national level.

\section{Model}

%We consider a downlink scenario, where photons go from the satellite to the ground stations, as the effects of the atmosphere are smoother compared to the uplink setup, where photons go from the ground stations to the satellite, \cite{andrews_laser_2005,gundogan2021}. We consider a single satellite link in a LEO orbit, at a height of around 350 (400) km, in a medium-inclination orbit, at a latitude of 42 $\degree$, going over the Iberian Peninsula, as depicted in Fig.\ref{fig:fig_1}(b). The satellite has an on-board entangled photon source probabilistically generating entangled photon pairs with correlated polarization.

\begin{figure*}[htp]
   \centering
    \includegraphics[width=\linewidth]{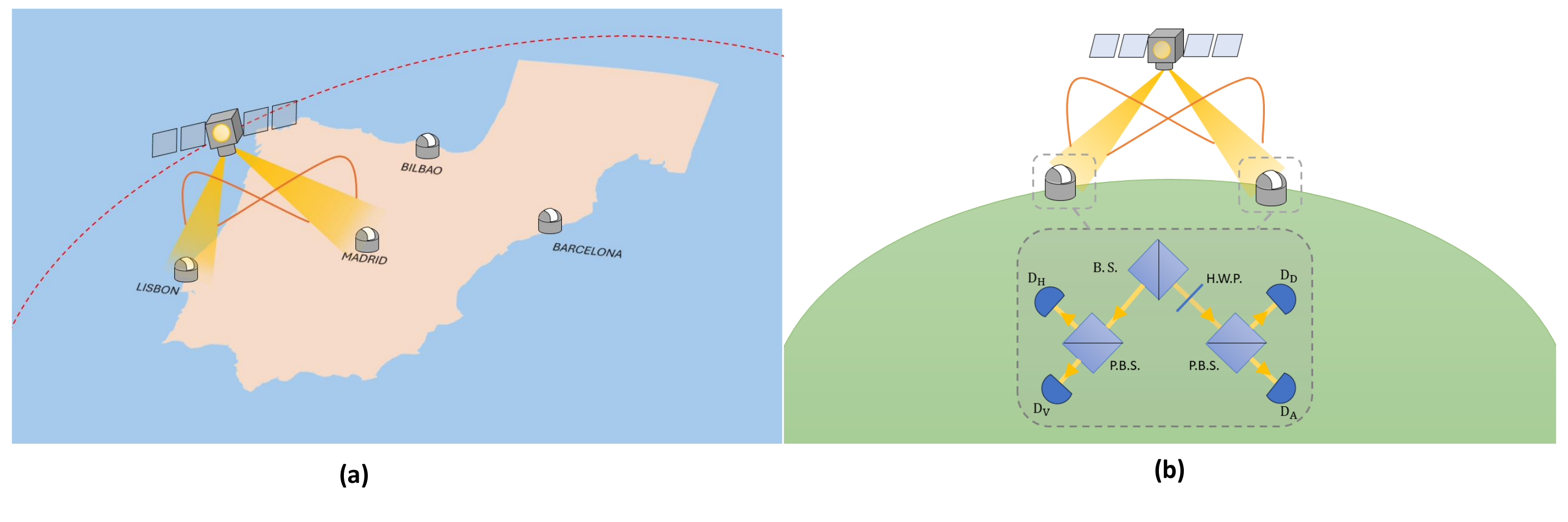}
    \caption{\textbf{Experimental setup. (a)} Ground stations of the regional quantum key distribution (QKD) network that we want to implement. The central node is Madrid, and the key distribution is always between Madrid and any of the other cities (Barcelona, Bilbao and Lisbon). \textbf{(b)} We consider a downlink scenario, with an entangled photon source firing photons from the satellite to the ground stations. At the ground stations, the photons are randomly measured in the X or Z basis. To perform such a measurement, the photons go through a 50/50 beam splitter (B.S.), followed by a half-wave plate (H.W.P.) in one of the outputs, which changes the basis of the input photons. Finally, before measuring, the photons go through a polarizing beam splitter (P.B.S.) such that we measure different polarizations- horizontal ($D_H$), vertical ($D_V$), diagonal ($D_D$) and antidiagonal ($D_A$)- at the output ports of the B.S. The measurements results are fed into a classical memory from which we can then extract secret keys for the two applications we consider.}
    \label{fig:fig_1}
\end{figure*}

We focus on a downlink configuration, where photons are transmitted from a satellite to ground stations. This setup is preferred because atmospheric disturbances are generally less severe than in the uplink case, where photons travel from the ground to the satellite \cite{andrews_laser_2005,gundogan2021}.  The satellite carries an onboard entangled photon source that probabilistically generates polarization-entangled photon pairs. Specifically, we consider a single satellite like Micius, with telescopes of radius of $15$ cm, a rate of entangled pairs per second of $5.9\cdot 10^{6}$ ph/s and pointing jitters of $\sim 10^{-6}$ rad \cite{yin_satellite-based_2017}, in a low Earth orbit (LEO) at an altitude of approximately 400 km, following a medium-inclination trajectory at a latitude of 42°, passing over the Iberian Peninsula, as illustrated in Fig.~\ref{fig:fig_1}(b).

We consider the BBM92 protocol \cite{yin_satellite-based_2017}, where the photons are measured randomly in either the Z or X basis at the ground stations. Only photon pairs detected in the same basis are retained in classical memories through postselection. We employ the BBM92 protocol, following the approach demonstrated by the Micius satellite \cite{yin_satellite-based_2017}. The figure of merit that we base our analysis on is the extracted key material; in general, the value is presented as a certain amount of \textbf{bits per second} (bps), but eventually it may also be convenient to also express it in terms of \textbf{bits per satellite pass}. The calculation involves three main steps, which are discussed in the Sections to follow: the model of the photonic qubit states generated by the entangled photon-pair source on-board the satellite; the effect of the optical channels between the satellite and the ground stations; and the effect of the measurement performed by the receivers in the presence of noise.

%At the ground stations, the photons are randomly measured in the Z or X basis, and then the ones measured on the same basis are postselected. We carry out the same protocol as Micius, the BBM92 protocol \cite{yin_satellite-based_2017}. 

\subsection{Entangled photon pair source}

A pump laser is used to drive a nonlinear optical crystal in both the clockwise and anticlockwise directions simultaneously, generating down-converted polarization-entangled photon pairs with orthogonal polarizations \cite{yin_entanglement-based_2020, Ma_2007}.

\begin{equation}
\begin{aligned}[b]
    |\psi_{\textrm{source}}\rangle \approx &(1-\lambda)\left(|0,0\rangle_{A,B}  \right.\\
    &+\left. \frac{\sqrt{2}\lambda^{1/2}}{(1-\lambda)^{1/2}}\frac{1}{\sqrt{2}}\left(|H\rangle_A|V\rangle_B + |V\rangle_A|H\rangle_B\right)  \right.\\
    &+\left.\frac{\sqrt{3}\lambda}{1-\lambda}\frac{1}{\sqrt{3}}\left(|2H\rangle_A|2V\rangle_B + |HV\rangle_A|HV\rangle_B \right.\right. \\
    &+ \left.\left.|2V\rangle_A|2H\rangle_B\right) + O\left(\sqrt{n+1}\frac{\lambda^{n/2}}{(1-\lambda)^{n/2}}\right)\right), 
    \label{eq:spdc_source}
\end{aligned}
\end{equation}

where the subscripts $A$ and $B$ correspond to two different spatial modes directed toward the ground stations of Alice and Bob respectively, $|H\rangle$ and $|V\rangle$ denote the horizontal and vertical polarization states, respectively. The state $|\Psi^{+}\rangle = 1/\sqrt{2}\left(|H\rangle_A|V\rangle_B + |V\rangle_A|H\rangle_B\right)$ denotes the desired state which corresponds to a maximally entangled state with anticorrelated polarization. 

The term $\mu=2\lambda$ is the average number of photon pairs generated per pulse, characterised by the brightness of the SPDC source \cite{Ma_2007}. The source operates in a weak-pump regime, where $\lambda \ll 1$, implying that the probability of emitting a photon pair is low. However, operating in this regime is necessary since it suppresses multi-photon events ($\propto |2H\rangle_A|2V\rangle_B + |HV\rangle_A|HV\rangle_B + |2V\rangle_A|2H\rangle_B$), which causes errors and lowers the secret key rate. In other words, there is trade-off between the rate and the fidelity of entanglement. In our simulation, we optimize $\lambda$ to obtain the highest SKR.

\subsection{Photon transmission}\label{sec_pT}

Photon loss during transmission to the two ground stations will change the received state from the one emitted by the SPDC source. We model this by passing the state in Eq.\eqref{eq:spdc_source} through fictitious beam splitters in each spatial mode, where one output of the beam splitters corresponds to the loss, while the other is the transmitted mode. In general, the two spatial modes experience different transmission probabilities due to the various distances from the satellite to each of the ground stations.

%By considering the different spatial modes in the transmission probability, we take into account the asymmetry of the channels due to the different satellite's trajectories over the ground stations. \mario{I would change the last sentence to: By considering the different spatial modes in the transmission probability, we take into account the asymmetry of the channels due to the different distances from the satellite to each of the ground stations.}

%\begin{eqnarray}
 %   \hat{a}^{\dagger}_{H(V)}|0\rangle_{A(B)} & \rightarrow & \sqrt{p_{T,A(B)}}\hat{a}^{\dagger}_{H(V)}|0\rangle_{A(B)} + \nonumber \\
  %  && \sqrt{1-p_{T,A(B)}}\hat{a}^{\dagger}_{H(V)}|0\rangle_{E_A,(E_B)},
   % \label{eq:BS_atmosphere}
%\end{eqnarray}

%where the subscripts $A$ and $B$ in the transmission probability distinguish the different paths the photons follow in the satellite-Alice(Bob) links, taking into account the asymmetry of the channels due to the satellite's trajectory over the ground stations.

\begin{figure}
    \centering
    \includegraphics[width=0.6\linewidth]{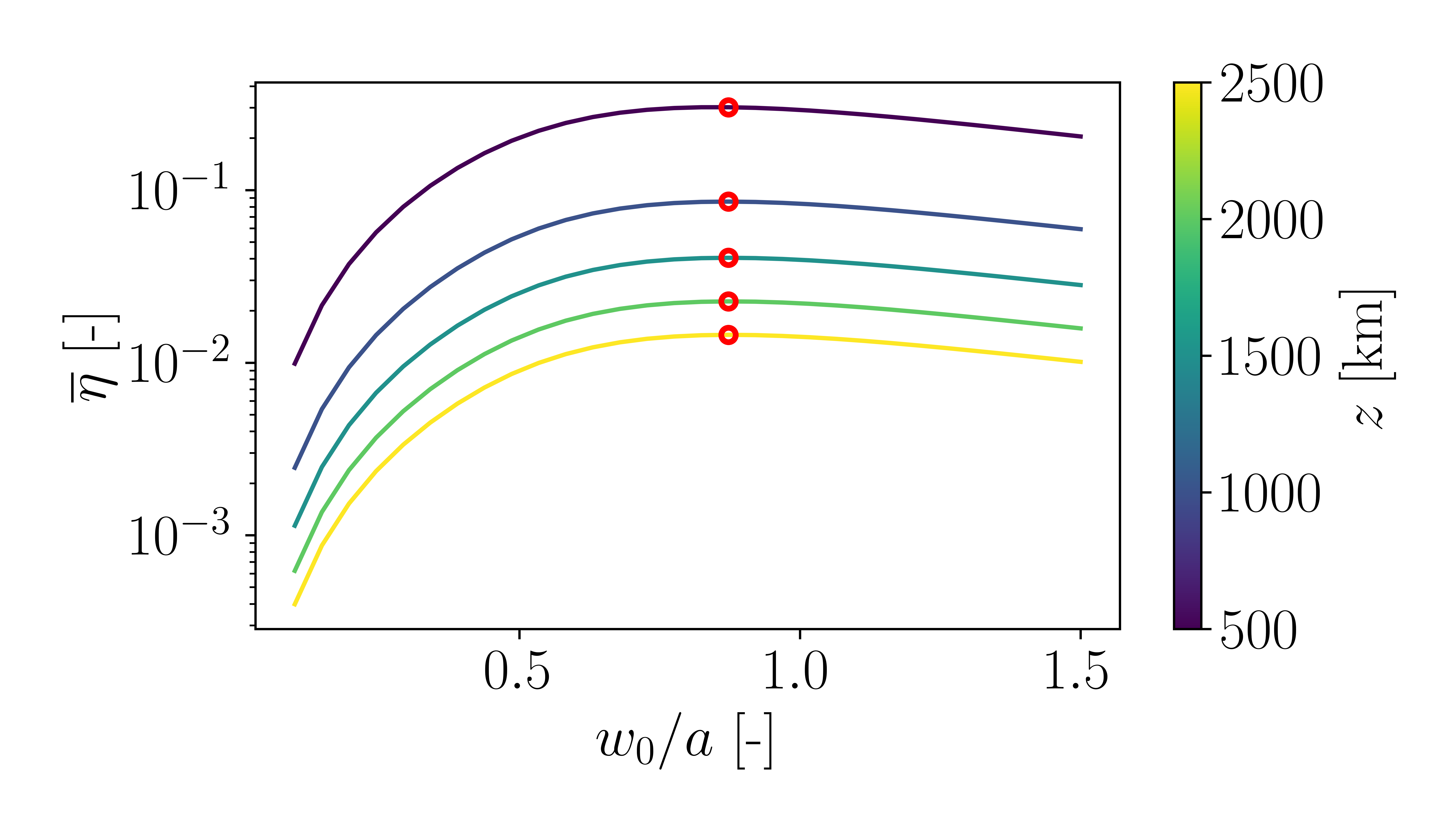}
    \caption{Optimization of the beamwaist for a pointing jitter $\sigma_\textrm{PJ}=0.47\;\mu\textrm{rad}$, satellite's aperture radius of $a=15\;\textrm{cm}$ and ground station's aperture radius of $60\;\textrm{cm}$. Different distances $z$ between the transmitter and the receiver are shown.}
    \label{fig_optw0}
\end{figure}

The free space transmission probability is estimated from the model in Ref. \cite{tubio2024satelliteassistedquantumcommunicationsingle}, which includes divergence, atmospheric absorption, and pointing jitter. For a more accurate estimation of the number of entangled photon-pairs captured by the ground station per second, an optical propagation model based on Ref. \cite{badas_seidel_2025}, which accounts for the truncation of the beam in the transmitter's telescope, is used (see Appendix \ref{app_B}). Previous works often assume a Gaussian beam where truncation of the beam in the transmitter can be neglected. Instead, we explicitly take this effect into account using the model from Ref. \cite{badas_seidel_2025}. This model enables the accurate evaluation of the average photon capture probability of each downlink channel, accounting for both the diffraction effects and the pointing jitter. By changing the beamwaist $w_0$ of the transmitted spatial Gaussian mode, the average photon capture probability $\overline{\eta}$ can be maximized. Fig.~\ref{fig_optw0} shows the optimization of the beamwaist for the apertures used in the present work. The results show that an optimum beamwaist can be found that maximizes the average photon capture probability (this maximum barely changes with the distance between the terminals). This maximum is the result of two counteracting effects when increasing the beamwaist. On the one hand, increasing the beamwaist increases the collimation of the beam and therefore results in a smaller spread of the spatial mode on the receiver aperture plane. On the other hand, as the beamwaist increases, the truncation effects due to the finite aperture of the transmitter start to be relevant. These truncation effects account for the telescope's transmission probability and the diffraction induced due to the edges of the aperture.

As a result of this optimization the spatial spread of the single photon mode in the ground station's aperture plane can be significantly reduced, compared to a case in which $w_0\ll a$ (where $a$ is the transmitter's telescope radius). This optimum beam, when combined with the pointing jitter statistics, results in a higher average photon capture probability $\overline{\eta}$. However, this effect comes along with an increase on the fluctuation of the photon capture probability $\eta$ around the average $\overline{\eta}$. Although these fluctuations can be very detrimental in classical communication links \cite{badas_seidel_2025}, this is not the case for quantum entanglement distribution. As shown in the Supplemental Material, the average photon capture probability is what matters on quantum entanglement distribution protocols affected by the pointing jitter stochastic process. These differences arise because classical systems decode information from large correlated bit strings, that require to have a stable power to get as many of these correlated bits as possible. Then classical error correction codes are applied to correct the bits flipped in the transmission. If the probability of being below a certain power threshold is high, the classical error correction codes can not compensate for the flipped bits, and information contained in the full bit string is lost. For quantum key distribution, however, we detect individual independent events on each receiver. Hence, in this case, the maximization of the number of photons captured in each receiver can be pursued.

The effect of the atmospheric channel in the performance of the system has also been modeled. The atmospheric attenuation and scattering model presented in \cite{Gustavo_paper} has been used. Regarding the atmospheric turbulence, under weak turbulence conditions, the effect of beam spread and wander has been considered negligible for the downlinks at limited zenith angles considered in this work \cite{andrews_laser_2005}. Moreover, the turbulence induced scintillation and the resulting coupling efficiency on the detector is accounted for on the coupling efficiency ($\textrm{CL}$) -- this efficiency is obtained assuming an adaptive optics system on the ground station \cite{chen_experimental_2015}. Furthermore, based on satellite-to-ground experimental observations reported in \cite{carrasco-casado_leo--ground_2016}, the depolarizing effect induced by atmospheric turbulence can be regarded as negligible.

Other dynamic atmospheric effects that significantly impact the quality of the QKD link depending on time of day and weather are also taken into account. This is possible due to the framework developed in \cite{Gustavo_paper}, where the sun irradiance -- used to calculate the stray light impinging on the single-photon detectors -- and atmospheric parameters -- such as cloud coverage factor, weather visibility, and the relative angle between the satellite and the ground station telescope -- are extracted at the ground station locations and used to determine noise levels and photon losses dynamically. We also model false detections assuming that there is a probability $p_{\textrm{dark}}$ that a detector clicks despite no photon being received. We assume that this event translates into a detected state with zero fidelity with respect to the desired entangled state, leading to a lower bound in the overall fidelity measured at the ground. 

\subsection{Photon measurement}

At the ground stations, the incoming photons are randomly measured in either the Z basis—defined by the states $|H\rangle$ (horizontal) and $|V\rangle$ (vertical)—or the X basis, defined by $|D\rangle = (|H\rangle + |V\rangle)/\sqrt{2}$ (diagonal) and $|A\rangle = (|H\rangle - |V\rangle)/\sqrt{2}$ (antidiagonal). To implement this random-basis measurement, the photons first pass through a 50/50 beam splitter (BS). At one of the BS output ports, a half-wave plate (HWP) is used to rotate the measurement basis from Z to X. Both outputs are then directed into a polarizing beam splitter (PBS), which spatially separates the polarization components, as illustrated in Fig.~\ref{fig:fig_1}(b).

Assuming the maximally-entangled state $|\Psi^{+}\rangle$ is received at the ground stations, it evolves as follows:
\begin{equation}
\begin{split}
    \frac{1}{\sqrt{2}}\big(|H\rangle_{A_a}|V\rangle_{B_a} + |V\rangle_{A_a} &|H\rangle_{B_a}\big)\xrightarrow{B.S.} \\
    \frac{1}{\sqrt{2}}\Bigg[&\frac{1}{\sqrt{2}}\left(|H\rangle_{A_c} + i|H\rangle_{A_d}\right)\frac{1}{\sqrt{2}}\left(|V\rangle_{B_c} + i|V\rangle_{B_d}\right) + \\
     &\frac{1}{\sqrt{2}}\left(|V\rangle_{A_c} + i|V\rangle_{A_d}\right)\frac{1}{\sqrt{2}}\left(|H\rangle_{B_c} + i|H\rangle_{B_d}\right)\Bigg].
    \label{eq:after_BS}
\end{split}
\end{equation}
Here, the subscripts $A_c$, $A_d$, $B_c$, and $B_d$ refer to the different spatial modes after the beam splitter for photons A and B, respectively. The output ports labelled “d” undergo a basis transformation via a HWP: $|H\rangle \rightarrow |D\rangle$ and $|V\rangle \rightarrow |A\rangle$.

Subsequently, each output passes through a PBS, which separates the polarization components before detection. The resulting state just before measurement is:
\begin{equation}
\begin{aligned}[b]
    |\psi_{\textrm{final}}\rangle = 
    \frac{1}{2\sqrt{2}}&\left[|H\rangle_{A_c}|0\rangle_{A_f}|0\rangle_{B_e}|V\rangle_{B_f} + i|H\rangle_{A_e}|0\rangle_{A_f}|0\rangle_{B_g}|A\rangle_{B_h}  \right. \\
    &+\left. i|D\rangle_{A_g}|0\rangle_{A_h}|0\rangle_{B_e}|V\rangle_{B_f} - |D\rangle_{A_g}|0\rangle_{A_h}|0\rangle_{B_g}|A\rangle_{B_h}  \right.\\
    &+ \left.|0\rangle_{A_e}|V\rangle_{A_f}|H\rangle_{B_e}|0\rangle_{B_f} + i|0\rangle_{A_e}|V\rangle_{A_f}|D\rangle_{B_g}|0\rangle_{B_h} \right. \\
   &+\left. i|0\rangle_{A_g}|A\rangle_{A_h}|H\rangle_{B_e}|0\rangle_{B_f} -  |0\rangle_{A_g}|A\rangle_{A_h}|D\rangle_{B_g}|0\rangle_{B_h}\right].
\end{aligned}
\label{eq:final_state}
\end{equation}

From Eq.~\eqref{eq:final_state}, we observe that in half of the cases, the photon pairs are measured in the same basis (either Z or X). Consequently, during post-selection, approximately 50\% of the events are discarded to ensure only measurements in matching bases are kept. The fidelity, $F$, of the final state, relevant to the calculations of secret key rate in the next section, is taken with respect to a maximally-entangled Bell state as (refer to Appendix B for the full calculation). %:
%\begin{equation}
%    F = \frac{\langle \Psi^{+} | \rho_{\textrm{final}}| \Psi^{+}\rangle}{Tr(\rho_{\textrm{final}})},
%    \label{eq:finalFid}
%\end{equation}
%where $\rho_{\textrm{final}} = |{\psi_{\textrm{final}}}\rangle\langle{\psi_{\textrm{final}}}|$, and $Tr\left(\cdot\right)$ denotes the trace operation.

\section{Performance}

\subsection{Secret Key Rate Calculation}

The computation of the SKR follows the model shown in \cite{Gustavo_paper}, where both the assumption of depolarizing channel and Werner states produced by the entangled photon-pair source have been used. This analysis provides a lower bound on the achievable secret key rate and takes as input the calculated fidelity of the state measured on the ground. Furthermore, the model enables one to split the effect of losses and decoherence (loss of fidelity) in the channel, calculating them separately only to combine them when calculating the final secret key rate expression:
\begin{equation}
    \textrm{SKR} = R_{\textrm{final}} \cdot \left( \textrm{max}\left(0, 1-\left(1+\xi\right)\mathcal{H}\left(\textrm{QBER}\right)\right) \right).
    \label{eq:SKR}
\end{equation}
In the expression above, $\mathcal{H}$ is Shannon's entropy function, $\xi$ is a factor representing the inneficiency of the classical protocol required for key extraction \cite{Ma_2007}, and the QBER is calculated based on the total state fidelity (see Appendix \ref{app_A}).
\begin{equation}
    \textrm{QBER} = \frac{1-\tfrac{3F+1}{4}}{2}.
\end{equation}

R$_{\textrm{final}}$ depends on the rate of the source in the satellite and the total channel efficiency, i.e., source-to-ground-station detector. This term can be decomposed into three main ones: the free-space path loss, governed by the divergence of the beam; the atmospheric attenuation factor, governed by the weather condition; and the detector efficiency, which varies significantly in case one chooses cryogenically-cooled detectors (higher efficiency, but fibre coupling is required), or free-space-coupled detectors (lower efficiency, no fibre coupling). For the purpose of this analysis, we choose to operate under the assumption that fibre coupling can be efficiently implemented via adaptive optics correction strategies, leading to an overall net improvement in the channel quality with the more efficient superconducting nanowire single-photon detectors (SNPDs). This assumption is also tied to a lower system jitter and lower detector dead times, which lead to higher signal-to-noise ratio (due to narrower available detection time windows and detection saturation rates).

In summary, the transmission from the satellite to the ground stations follows the expression
\begin{align}
    p_{T,A} = &\overline{\eta}_{Sat-GndA} \;\textrm{ATML}_{Sat-GndA} \;\textrm{CL}_{Sat-GndA}\\
    p_{T,B} = &\overline{\eta}_{Sat-GndB} \;\textrm{ATML}_{Sat-GndB} \;\textrm{CL}_{Sat-GndB}.
\end{align}
$\overline{\eta}$ accounts for the diffraction and pointing jitter losses (see Section~\ref{sec_pT}), ATML for atmospheric attenuation and scattering loss, CL for coupling loss, and GndA/GndB for the two ground stations of interest. The detailed description of how each of these terms is evaluated can be found in the Supplemental Material. The reason for evaluating the two paths individually is the channel asymmetry during satellite transit, not only in terms of the dynamic conditions (such as weather conditions) but also the distance between the spacecraft and either ground station.

Equipped with the overall loss term, the probabilities -- per attempt -- associated with the detection of a valid pair can be translated into raw detection rates via the source pair emission rate R. These are further translated into expected secret key rate using a key extraction algorithm that implements: basis reconciliation, error correction, and privacy amplification. As mentioned before, depending on the channel quality, the raw detection material that must be spent in order to produce a final key will change according to the classical protocols being used. Since this procedure can be quite computationally intensive and would have to be executed for each time step at each satellite pass, a look-up table was generated by feeding the software with a grid of parameters (QBER and block length) -- refer to Appendix C for discussions and results.

\subsection{Representative System Evaluation}

For the course of one month, specifically June 2025, we compute the SKR per satellite pass that reached the ground stations, while accounting for real-time weather conditions. Our analysis focuses on three links: Madrid–Barcelona, Madrid–Bilbao, and Madrid–Lisbon. We begin by computing the SKR per pass and, based on this, derive the usable secret key material per pass. By averaging the usable key material per pass and taking into account the percentage of successful passes, i.e. the passes where communication between the satellite and the ground stations is established, we obtain the average secret key material rate, which we then compare against the requirements of Telefónica and JPMorgan use cases.

%To evaluate the protocol’s performance and quantify its optimization, we compute the SKR per satellite pass that reaches the ground stations over the course of one month—specifically, April 2025—while accounting for real-time weather conditions. Our analysis focuses on three links: Madrid–Barcelona, Madrid–Bilbao, and Madrid–Lisbon. We begin by computing the SKR per pass and, based on this, derive the usable secret key material per pass (see section \ref{sec:skr}). By averaging the usable key material per pass and taking into account the percentage of successful passes, we obtain the average secret key material rate, which we then compare against the requirements of Telefónica and JPMorgan use cases.

%\begin{figure*}[htp]
 %  \centering
  %  \includegraphics[width=0.95\linewidth]{results_Mad_Bilbo_new.pdf}
   % \caption{\textbf{ Secret key material between Madrid and Bilbo} The blue data shows the usable key without carrying out the beam optimization, and the red data shows the results when applying the optimization of the beam. The discontinuous red line shows the average key per pass.}
   % \label{fig:results}
%\end{figure*}

\begin{figure*}[htp]
   \centering
    \includegraphics[width=\linewidth]{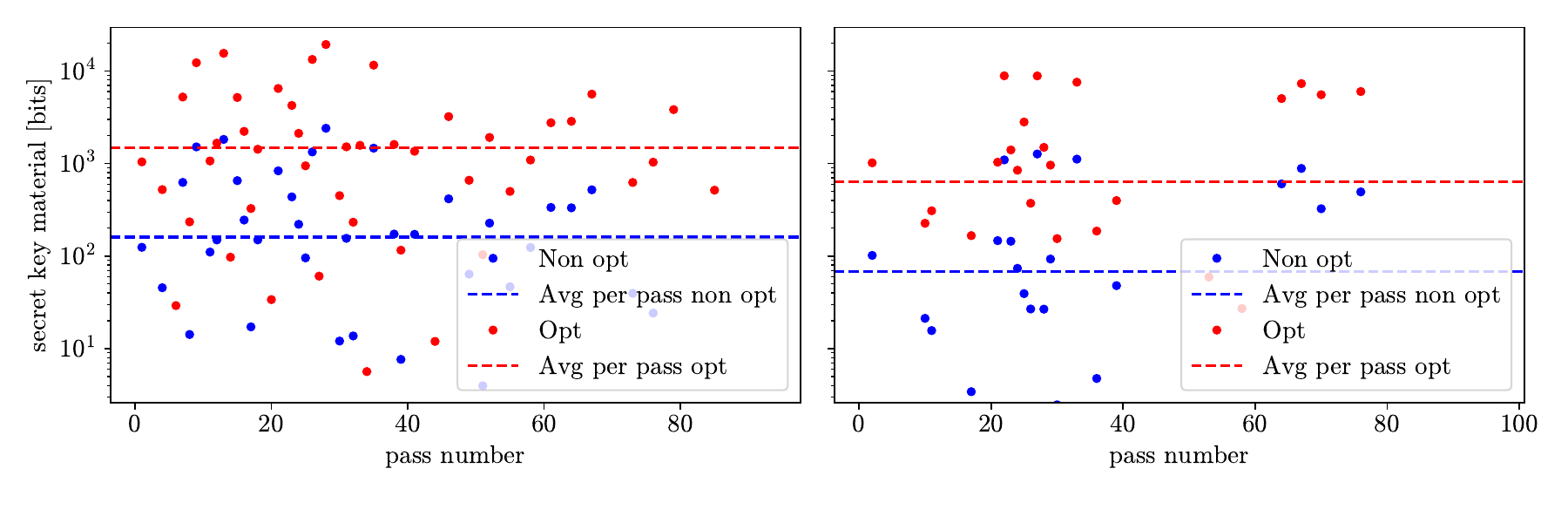}
    \caption{\textbf{ Secret key material between Madrid and Barcelona (left), and Madrid and Bilbao (right).} The blue data shows the usable key without carrying out the beam optimization, and the red data shows the results when applying the optimization of the beam. The discontinuous red line shows the average key per pass. In the link Madrid-Barcelona, the successful passes, where key is shared between the satellite and the ground stations, are around 39\% for the non-optimized case and 48\% for the optimized case. While in the Madrid-Bilbao link, the successful passes for the non-optimized case are around 22\% and for the optimized case 24\%, leading to a lower average of secret key material.}
    \label{fig:results}
\end{figure*}

In Fig.\ref{fig:results}, we present the secret key material per satellite pass and its average for both the non-optimized and optimized cases on the Madrid Barcelona and Madrid–Bilbao links. The figure shows that real-time beam optimization results in an increase of the average secret key material by an order of magnitude which increases the SKR to a point where enough secret bits are produced to run the applications.

%In Fig.\ref{fig:results}, we show the secret key material per pass and its average for the non optimized and for the optimized case in the Madrid-Bilbao link. From the aforementioned figure, we see that the optimization of the beam in real time shows an increase of the average secret key material by a factor of ten. 

%\begin{table*}[htp]
%\begin{tabular}{|l|l|l|l|l|l|l|}
%\toprule
%   Links &  Non opt [keys/s] & Telefonica & JPMorgan & Opt [keys/s] & Telefonica & JPMorgan \\
%\midrule
% Mad-Bilb &     0.29 & Yes & No &2.44 &  Yes & Yes  \\
%Mad-Bcn &     0.12 & Yes & No & 1.05 &  Yes & No  \\
% Mad-Lis &     0.29 & Yes & No &  2.62 &  Yes &  Yes  \\
%\bottomrule
%\end{tabular}
%\caption{Results for an entangled photon source like Micius $0.5\cdot 10^{7}$ Hz.}
%\label{tab:micius_results}
%\end{table*}

%\begin{table*}[htp]
%\begin{tabular}{|l|l|l|l|l|l|l|}
%\toprule
 %  Links &  Non opt [keys/s] & Telefonica & JPMorgan & Opt [keys/s] & Telefonica & JPMorgan \\
%\midrule
 %Mad-Bilb &     0.003 & No & No &0.018 &  Yes & No  \\
%Mad-Bcn &     0.002 & No & No & 0.012 &  Yes & No  \\
 %Mad-Lis &     0.002 & No & No &  0.018 &  Yes &  No  \\
%\bottomrule
%\end{tabular}
%\caption{Results for an entangled photon source like Micius $5.9\cdot 10^{6}$ Hz\cite{yin_satellite-based_2017}.}
%\label{tab:micius_results}
%\end{table*}

\begin{table*}[htp]
\resizebox{\textwidth}{!}{
    \begin{tabular}{|l|l|l|l|l|l|l|}
        \hline
           Links &  Non opt [keys/s] & Telefonica & JPMorgan & Opt [keys/s] & Telefonica & JPMorgan \\ \hline 
         Mad-Bilb &     0.003 & No & No &0.028 &  Yes & No  \\ \hline
        Mad-Bcn &     0.012 &   Yes & No & 0.132 &  Yes & No  \\ \hline
         Mad-Lis &     0.004 & No & No &  0.055 &  Yes &  No  \\ \hline
    \end{tabular}}
    \caption{Results for an entangled photon source like Micius $5.9\cdot 10^{6}$ Hz\cite{yin_satellite-based_2017}.}
    \label{tab:micius_results}
\end{table*}

Table \ref{tab:micius_results} summarizes the computed results for the different links and compares them against two use cases: Telefónica, which requires 256 keys every 12 hours (equivalent to 0.006 keys/s), and JPMorgan, which requires 256 keys every 2 minutes (2.13 keys/s). We observe that the 12-hour key renewal, used for communication between hospitals, is achieved across all links when the beam optimization is carried out. We do also see that for the link Madrid-Barcelona, this last use case can also be implemented without the optimization. This is due to the high percentage of successful passes, i.e., the key is shared between the satellite and the ground stations, we observe in that link during the month we are evaluating: around 40\% for the non-optimized case and 50\% for the optimized. This contrasts with the other two links, where the percentage of successful passes is around 20–25\%, and up to 34\% for the optimized case between Madrid and Lisbon. The passes where the communication is established between the satellite and the ground station are higher in the optimized case due to the fact that in those cases the overall channel efficiency improves; therefore, there will be cases where the SKR is zero in the non optimized cases and non-zero in the optimized case.

However, for the more demanding use case of QKD-based VPNs, where the key must be refreshed every 2 minutes, the required key rate is not met. To get to such key requirements, we would need to improve the entangled photon source rate to the order of 1 GHz \cite{merolla2022high} (for more details check Appendix \ref{app_D}). Additionally, we could also boost the SKR by not only increasing the photon source rate, but also combining it with the decrease in coupling loss, $CL$, which we consider 0.5 due to the scintillation effect of the lowest parts of the atmosphere \cite{chen_experimental_2015,marulanda_acosta_analysis_2024,gruneisen_modeling_2017}. This coupling efficiency could be increased by considering free space coupling into the single photon detectors \cite{mueller_free-space_2021}, or multimode fiber coupling \cite{zheng_free-space_2016,trinh_experimental_2023}. 
For distances beyond the Iberian Peninsula, constellations of satellites could be used.

%In table \ref{tab:micius_results}, we compare the results computed for the different links and compare them for the use cases of Telefónica, $256 \textrm{ keys}/12 \textrm{ hours}= 0.006 \textrm{ keys/s}$, and JPMorgan, $256 \textrm{ keys}/ 120 \textrm{ s}= 2.13 \textrm{ keys/s}$. We see that the renewal of the key each 12 hours done in the communication between hospitals is achieved for every link and also in the worst case scenario, without the optimization. For the most demanding case, when implementing QKD-based VPNs, where the key is renewed each 2 minutes, we see that we get to those values of key rate just when implementing the optimization for the link between Madrid and Bilbao, and Madrid and Lisbon. We do not get the desired key rate for the Madrid-Barcelona link due to the trajectory of the satellite, which passes more far away from Barcelona compared to the rest of the ground stations.

\section{Conclusion}

In summary, we have demonstrated how single-satellite architectures enable the extension of quantum key distribution (QKD) from metropolitan-scale deployments \cite{telefonica_vithas_qkd_2025} to national-scale distances. Our approach employs an entanglement-based QKD protocol, which eliminates the need for trusted nodes in the key distribution process, thereby reducing the risk of eavesdropping. Furthermore, we introduced a beam optimization method that accounts for satellite pointing jitter and the truncation effects due to the finite size of the transmitter's aperture, maximizing the transmission probability. This optimization allows us to meet the key rate requirement for the current use cases of secure communication between two hospitals \cite{telefonica_vithas_qkd_2025}, using state-of-the-art satellite technology \cite{yin_satellite-based_2017}. Furthermore, by upgrading the rate of the entangled photon source to 1GHz, which can be reached with current technology \cite{merolla2022high}, we could also meet the requirement for the most demanding use case \cite{VPN_JPMorgan}.

%In summary, we have shown how single-satellite-based architectures allow to go from the metropolitan distribution of quantum key \cite{VPN_JPMorgan,telefonica_vithas_qkd_2025} to national distances. Additionally, we show a beam optimization which takes into account the pointing jitter of the satellite and the distance of the satellite from the ground station to get the maximum possible transmission probability. Such optimization, allows us to meet the key rate requirements for even the most demanding use case for a state-of-the-art satellite \cite{yin_satellite-based_2017}.

In those use cases, hybrid quantum algorithms are employed, where only the first layer involves quantum processing. As a natural progression, future work should investigate fully quantum algorithms and their corresponding key rate requirements.

%By upgrading the rate of the entangled photon source \textbf{[add references]}, the system can be adapted in two ways: placing the satellite in higher orbits to increase communication windows, and (or) employing smaller, more cost-effective satellites

%We use an entanglement-based QKD protocol, avoiding the use of trusted nodes in the distribution of keys, decreasing the risk of being eavesdropped.

%By upgrading the entangled photon source rate \textbf{add references}, we can move the satellite to higher orbits, allowing larger communication windows, or to use smaller satellites and consequently cheaper.

Additionally, in our current setup, we assume trusted ground stations for classical information storage. Incorporating quantum memories at the ground level would allow for the use of untrusted ground stations and enable the distribution of entanglement across nodes. This would open the door to a broader range of quantum technologies beyond QKD, including blind quantum computing \cite{Fitzsimons2017PrivateProtocols,VanMeter2016TheComputing} and quantum-enhanced sensing networks \cite{Komar2014,Guo2020,Liu2021}.

%To extend quantum communication to larger distances, the integration of quantum memories onboard satellites is needed \cite{tubio2024satelliteassistedquantumcommunicationsingle,Gustavo_paper,Wallnofer_2022,Liorni_2021}. Currently, we assume trusted ground stations for classical information storage. However, incorporating quantum memories at the ground level would allow for the use of untrusted ground stations and enable the distribution of entanglement across nodes. This would open the door to a broader range of quantum technologies beyond QKD, including blind quantum computing \cite{Fitzsimons2017PrivateProtocols,VanMeter2016TheComputing} and quantum-enhanced sensing networks \cite{Komar2014,Guo2020,Liu2021}.

%The next step to achieve larger distances, is the use of quantum memories on board. \textbf{citations}. Additionally, we are assuming trusted ground stations where we store the information classically. The use of quantum memories on ground allows us to have untrusted ground stations and to distribute entanglement between nodes, which we can use for other applications besides QKD, such as blind quantum computing or enhanced sensing networks. 

%In current use cases, we use hybrid algorithms, where just the first layer of it, it is quantum. As a next step, it would be interesting to study completely quantum algorithms and the key requirements for those cases.

\begin{backmatter}
\bmsection{Funding}
% Content in the funding section will be generated entirely from details submitted to Prism. Authors may add placeholder text in this section to assess length, but any text added to this section will be replaced during production and will display official funder names along with any grant numbers provided. If additional details about a funder are required, they may be added to the Acknowledgment, even if this duplicates some information in the funding section. For preprint submissions, please include funder names and grant numbers in the manuscript.

\bmsection{Acknowledgment}
V.D.T. acknowledges helpful discussions with Antonio Agustín Pastor Perales and María Ruiz Fernández de Arcaya. V.D.T. and J.B. acknowledge funding from the NWO Gravitation Program Quantum Software Consortium (Project QSC No. 024.003.037). J.B. acknowledges support from The AWS Quantum Discovery Fund at the Harvard Quantum Initiative. M.B. acknowledges the funding from the NWO Perspectief FREE Program (P19-13).

\bmsection{Disclosures}
The authors declare no conflicts of interest.

\bmsection{Data availability} The data of the results of this paper are openly available on 4TU.ResearchData: "Data underlying the publication "Quantum Key Distribution in the Iberian Peninsula", at \url{https://doi.org/10.4121/eda0897b-8157-4ab3-9a1b-db10d1baf33e.v1}, Ref.\cite{dataset}.

\bmsection{Supplemental document}
See Supplement 1 for supporting content.

\end{backmatter}

\appendix

\section{Free space channel model} \label{app_B}
In this appendix, we discuss the details of our free space channel model. This model considers the effects of the beam propagation and the pointing jitter in a coupled manner. Furthermore, as the maximum average photon capture probability is obtained when the spatial Gaussian mode describing the single photon in the transmitter aperture is highly truncated, the model presented in this appendix also accounts for the diffraction effects due to this truncation. A more detailed explanation of this model can be found in \cite{badas_seidel_2025}.

The beamwaist $w_0$ of the spatial Gaussian mode is located at the satellite's telescope aperture. The telescope on board the satellite has a radius $a$. The field at the ground station's aperture plane, located at a distance $z$, can be computed numerically through Fresnel propagation \cite{goodman_introduction_2005}. This propagation gives the probability density function of the spatial location of the single photon, i.e. $I(x,y)$, in the ground station's aperture plane. Then, a convolution can be performed to compute the probability of capturing the single photon $g(x_0,y_0)$ when the center of the beam is displaced to $(x_0, y_0)$ in the ground station's reference system. This convolution can be computed using the convolution theorem by
\begin{equation}
    g(x_0,y_0) = \mathcal{F}^{-1}{\mathcal{F}(I)\cdot\mathcal{F}(A)}
\end{equation}
where $A(x,y)$ is the aperture function of the ground station's telescope and $\mathcal{F}$ is the bidimensional Fourier transform. The satellite pointing jitter is modeled as a bidimensional Gaussian probability density function of the location of the center of the beam at the ground station frame of reference $(x_0, y_0)$, 
\begin{equation}
    f_\textrm{PJ}(x_0, y_0) = \dfrac{1}{2\sigma^2}\exp\left(-\dfrac{x_0^2+y_0^2}{2\sigma^2}\right)
\end{equation}
where $\sigma=z\sigma_\textrm{PJ}$, and $\sigma_\textrm{PJ}$ is the angular pointing jitter. By combining the previous equations, the probability density function of the photon capture probability can be computed as \cite{rohatgi_introduction_2001}
\begin{equation}
    f(\eta) = \iint_{\mathbb{R}^2}f_\textrm{PJ}(x_0, y_0)\;\delta[\eta-g(x_0,y_0)]\;dx_0\,dy_0
\end{equation}
where $\delta(x)$ is the Dirac delta function. Finally, the average of this probability density function gives the average photon capture probability. As shown in the Supplementary Material, the latter is to be maximized in the context of quantum entanglement based QKD for optimal performance of the system. This maximization is done by varying the beamwaist $w_0$ of the spatial Gaussian mode transmitted.

\section{Fidelity of the final state} \label{app_A}

In this appendix, we provide a more detailed calculation of how we compute the fidelity required for the secret key rate computation. The final state is of the form:
\begin{equation}
    \rho_{\textrm{final}} = A|\psi\rangle\langle \psi| + B\rho_{\textrm{deph}} + C\rho_{\textrm{garb}},
\end{equation}

where $|\psi\rangle$ is the desired target state which gives fidelity 1, $\rho_{\textrm{deph}}$ are the multi-photon states, where one photon for each elementary link has been lost, resulting in a dephased states which gives fidelity smaller than one, and $\rho_{\textrm{garb}}$ is the non-specified 'garbage' density matrix with zero overlap with the desired target state. The (normalized) density matrix of the dephased state is the following:

\begin{equation}
\begin{aligned}[b]
\rho_{\textrm{deph}} =\dfrac{1}{12}\left(5|H\rangle|V\rangle\langle H|\langle V| + 5|V\rangle|H\rangle\langle V|\langle H| + |H\rangle|H\rangle\langle H|\langle H| + |V\rangle|V\rangle\langle V|\langle V|\right),
\end{aligned}
\label{eq:dephased_state}
\end{equation}

being the fidelity of the previous state:

\begin{equation}
    F_{\textrm{deph}} = \langle \psi | \rho_{\textrm{deph}} | \psi \rangle = 5/12.
\end{equation}

Therefore, the overall fidelity is:

\begin{equation}
    F = \frac{\langle \psi | \rho_{\textrm{final}}| \psi\rangle}{Tr(\rho_{\textrm{final}})} = \frac{A + \frac{5}{12}B}{A + B + C},
\end{equation}

where the parameters $A,B,C$ are the following:

\begin{eqnarray}
A & = & \lambda(1-\lambda)p_{\textrm{T,A}}p_{\textrm{T,A}}, \\
B & = & 5 \lambda^2p_{\textrm{T,A}}p_{\textrm{T,B}}(1-p_{\textrm{T,A}})(1-p_{\textrm{T,B}}),  \\
C & = & \lambda(1-\lambda) \left(p_{\textrm{T,A}}(1-p_{\textrm{T,B}})p_{\textrm{dark}} + p_{\textrm{T,B}}(1-p_{\textrm{T,A}})p_{\textrm{dark}}  \right.\nonumber \\
&& + \left. (1-p_{\textrm{T,A}})(1-p_{\textrm{T,B}})p_{\textrm{dark}}^2\right)  \nonumber \\
&& + \lambda^2 \left(3p_{\textrm{T,A}}(1-p_{\textrm{T,A}})(1-p_{\textrm{T,B}})^2p_{\textrm{dark}}  \right. \nonumber \\
&&\left. +3p_{\textrm{T,B}}(1-p_{\textrm{T,B}})(1-p_{\textrm{T,A}})^2p_{\textrm{dark}} \right.\nonumber \\
&& \left. + \frac{3}{2}(1-p_{\textrm{T,A}})^2(1-p_{\textrm{T,B}})^2p_{\textrm{dark}}^2\right) \nonumber \\
&&+\frac{1}{2}(1-\lambda)^2p_{\textrm{dark}}^2.
\end{eqnarray}

\section{Key Extraction}

The key extraction protocol is performed after the quantum state transmission and reception phase of QKD, leading to a so-called block of detection events. The two communicating parties in the BBM92 configuration transmit the time tagged information about the (in this case, passively) chosen measurement bases. Basis reconciliation and time filtering allow both parties to discard all events that correspond to mismatched bases or those that fall outside the time window of acceptance. At this point, the protocol requires the parties to sacrifice a subset of the remaining events by disclosing the detection outcome in order to calculate the QBER. If the value is below the threshold for security, the protocol moves on to the next stage. Depending on the value of the QBER, however, the amount of privacy amplification rounds will vary: for low values of QBER, the expected mutual information between the communicating parties and a potential adversary is low, indicating that the security of the key is high; if the value of the QBER approaches the threshold, more rounds are necessary to reduce the mutual information and enforce security of the key material. 

\begin{figure*}[htp]
   \centering
    \includegraphics[width=\linewidth]{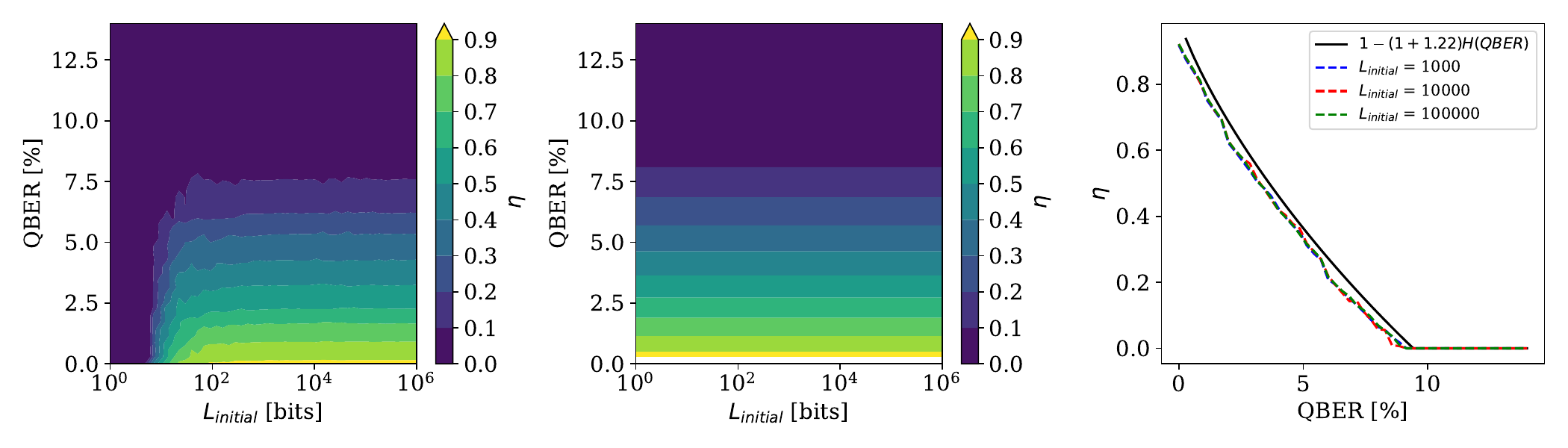}
    \caption{\textbf{Key Extraction Efficiency - Look Up Table} The heatmap shows the value of $\eta_{LUT}$ (Eq. \ref{eq:SKRLUT}) as a function of the block length and the estimated QBER.}
    \label{fig:LUT}
\end{figure*}

The factor on the righthand side of Eq. \ref{eq:SKR} of the Main Text indicates the ratio between the amount of raw detection events before privacy amplification and error correction rounds and the final amount of secure bits output by the protocol. Practical operation of QKD systems requires online computation of the QBER and execution of the subsequent rounds of classical communication that enable producing a key. The complexity of the steps, and the fact that they may be executed several times, lead to the development of dedicated hardware to tackle this problem efficiently and ensuring that it does not represent a bottleneck in the key distribution link. For the analysis presented in this work, several satellite passes have been evaluated over multiple days, which translates into a heavy computational task to execute the classical protocol for each block of raw detection events. Therefore, a look-up table (LUT) was created allowing one to efficiently and accurately estimating the value of the so-called \textbf{key extraction efficiency} ($\eta_{LUT}$) given a block length and the corresponding QBER value. In other words:
\begin{equation}\begin{split}
    \textrm{SKR} &= R_{\textrm{final}} \cdot \left( \textrm{max}\left(0, 1-\left(1+\xi\right)\mathcal{H}\left(\textrm{QBER}\right)\right) \right)\\
    &=R_{\textrm{final}} \cdot \eta_{LUT}\left[L_{block};QBER\right].
    \label{eq:SKRLUT}
\end{split}\end{equation}
The look-up table was generated by artificially inputting bit strings -- of different length and error rate -- to a key extraction software implementing the Cascade error correction protocol \cite{attema2021optimizing, github_keyExtraction}. For each combination of $\left[L_{block}; QBER\right]$, $\eta_{LUT}$ -- presented as a heatmap in Figure \ref{fig:LUT}-a -- outputs a value of that can be immediately used to calculate the expected key material generated by the QKD protocol. In Figure \ref{fig:LUT}-b, we present the same calculations but considering a static value of $\xi=1.22$, which approximates the realistic results for large block lengths ($L\geq 10^4$). Due to the limited availability of the satellite during the pass, and the dynamic weather conditions that can significantly diminish the key material generated on the ground, the regime of $L\leq10^4$ is relevant for the analysis of the results and justifies the use of $\eta_{LUT}$.

\section{Results with an entangled pair photon source rate of $10^{9}$ Hz} \label{app_D}

In this appendix, we provide the results we get with an entangled photon source rate 3 orders of magnitude higher than Micius. As we see in Table \ref{tab:results_10_9}, we always get the key requirements for Telefónica´s use case, and with the optimization with get the more demanding use case of refreshing the key each two minutes from JPMorgan.

\begin{table*}[htp]
\resizebox{\textwidth}{!}{
    \begin{tabular}{|l|l|l|l|l|l|l|}
        \hline
           Links &  Non opt [keys/s] & Telefonica & JPMorgan & Opt [keys/s] & Telefonica & JPMorgan \\ \hline
         Mad-Bilb &     0.565 & Yes & No & 5.005 &  Yes & Yes  \\ \hline
        Mad-Bcn &     2.367 & Yes & Yes & 22.978 &  Yes & Yes  \\ \hline
         Mad-Lis &     1.031 & Yes & No &  10.378 &  Yes &  Yes  \\ \hline
    \end{tabular}}
    \caption{Results for an entangled photon source of $10^{9}$ Hz.}
    \label{tab:results_10_9}
\end{table*}

\bibliography{main.bib}

\end{document}